# The Digital Omnibus on AI, Legislative Legitimacy and the Dynamics of AI Regulation

Donal Casey* (Uppsala University) and Liane Colonna* (Stockholm University)

**Abstract:** Driving the Digital Omnibus on AI are growing concerns within the European Union about economic growth, competitiveness, innovation and regulatory simplification. What is particularly striking about the Digital Omnibus on AI is that it seeks to amend the AI Act that entered into force less than two years ago in August 2024. This raises the question of how we can understand both the need and urgency with which the European institutions have sought amend a centrepiece of EU legislation that is still in its infancy.

In addressing this question, we examine the Digital Omnibus on AI through the lens of legislative legitimacy. In so doing, we ask how the specific dynamics of AI have shaped the legitimacy of AI regulation. Here, we are concerned not only with the issue of technological change, but with the broader dynamics of the evolving 'socio-technical landscape' of AI. In addressing this question, we argue that three specific dynamics have created a legitimacy dilemma for the EU institutions in the context of the AI Act: the race for AI regulation; the race for AI dominance and; the race for regulatory connection. As a response and proposed solution to this legitimacy dilemma, we further contend that, as a solution to the legitimacy dilemma caused by these three dynamics, the Digital Omnibus on AI in turn reshapes the legitimacy of the AI Act as it prioritises political and operational rationalities over legal and cultural rationalities.

## Introduction

The Digital Omnibus on AI, part of the Digital Omnibus package, was introduced by the European Commission in November 2025 and agreement on a final text was reached with the European Parliament and the Council of the European Union in May 2026.[1] Driving the Digital Omnibus on AI, and indeed the Digital Omnibus Package, are growing concerns within the European Union about economic growth, competitiveness, innovation and regulatory simplification. What is particularly striking about the Digital Omnibus on AI is that it seeks to amend the AI Act that entered into force less than two years ago in August 2024. This raises the question of how we can understand both the need and urgency with which the European institutions have sought amend a centrepiece of EU legislation that is still in its infancy.

In addressing this question, we examine the Digital Omnibus on AI through the lens of legislative legitimacy. In so doing, we ask how the specific dynamics of AI have shaped the legitimacy of AI regulation. Here, we are concerned not only with the issue of technological change, but with the broader dynamics of the evolving 'socio-technical landscape' of AI.[2] In addressing this question, we argue that three specific dynamics have created a legitimacy dilemma for the EU institutions in the context of the AI Act: the race for AI regulation;[3] the race for AI dominance and;[4] the race for regulatory connection (the pacing problem).[5] As a response and proposed solution to this legitimacy dilemma, we further contend that, as a solution to the legitimacy dilemma caused by these three dynamics, the Digital Omnibus on AI in turn reshapes the legitimacy of the AI Act as it prioritises political and operational rationalities over legal and cultural rationalities. We develop this argument in three parts.

We begin our analysis by setting out how we conceptualise legislative legitimacy. In so doing, we draw upon Wahlgren's theory of legislative legitimacy and bring it into conversation with wider discussions on the legitimacy of regulation.[6] In his work, Wahlgren sets out a framework of legislative legitimacy that rests upon five rationalities, namely political, legal, cultural, operational and internal rationalities.[7] Whilst this framework sets out predetermined 'standards

* This work was partially supported by the Wallenberg AI, Autonomous Systems and Software Program – Humanity and Society (WASP-HS) funded by the Marianne and Marcus Wallenberg Foundation and the Marcus and Amalia Wallenberg Foundation.

[1] European Commission, 'Proposal for a Regulation of the European Parliament and of the Council Amending Regulations (EU) 2024/1689 and (EU) 2018/1139 as Regards the Simplification of the Implementation of Harmonised Rules on Artificial Intelligence (Digital Omnibus on AI)' COM(2025) 836 final; Council of the European Union, 'Proposal for a Regulation of the European Parliament and of the Council Amending Regulations (EU) 2024/1689 and (EU) 2018/1139 as Regards the Simplification of the Implementation of Harmonised Rules on Artificial Intelligence - Letter Sent to the European Parliament' (2026) Information Note ST 9247 2026 INIT.

[2] Lyria Bennett Moses, 'How to Think about Law, Regulation and Technology: Problems with "Technology" as a Regulatory Target' (2013) 5 Law, Innovation and Technology 1.

[3] Nathalie A Smuha, 'From a "Race to AI" to a "Race to AI Regulation": Regulatory Competition for Artificial Intelligence' (2021) 13 Law, Innovation and Technology 57.

[4] Daniel Mügge, Regine Paul and Vali Stan, *The AI Matrix: Profits, Power, Politics* (Agenda Publishing 2026).

[5] Roger Brownsword, *Rights, Regulation, and the Technological Revolution* (Oxford University Press 2008).

[6] Peter Wahlgren, 'The Legitimacy Sphere: Between Law, Culture, Politics and Enforceability' (2010) 56 Scandinavian Studies in Law 428.

[7] Id.

and criteria of legitimacy' at its heart it is also a concern with congruence between legislation, political objectives and broadly held societal beliefs, expectations and interests.[8]

Building upon our understanding of legitimacy, we argue in Section 2 that three dynamics connected to the rapidly evolving socio-technical landscape of AI have created a legitimacy dilemma for the EU institutions in the context of the AI Act. The first dynamic is the race for AI regulation.[9] We argue that, in the haste to regulate AI, the EU institutions have created a complex system of legislative rules without ensuring that the regulatory infrastructure of standards, guidelines and oversight mechanisms are in place. This has resulted in a legitimacy dilemma driven by concerns about the actual implementation and operation of the legislative requirements. The second dynamic is the race to AI dominance, which has called into question the approach of the AI Act that on its face sought to prioritise rights over innovation. The race to AI dominance creates a legitimacy dilemma driven by changing political priorities, but also a sense of urgency that amplifies such a dilemma for the EU institutions. The final dynamic, the race for regulatory connection, is driven by the inevitable mismatch between an evolving socio-technical landscape and the existing regulatory regime, either in relation to the changing nature of technology, the changing ways in which technology is used or the changing social values as they relates to a technology.[10] The race for regulatory connection creates a legitimacy dilemma as it puts the functional, political and cultural basis of legislative legitimacy in direct conflict with the legal rationality's concerns with stability, predictability and certainty.

In Section 3, we examine the Digital Omnibus on AI as a response and proposed solution to the legitimacy dilemma through Wahlgren's framework of legislative legitimacy.[11] In so doing, we argue that whilst the Digital Omnibus on AI sought to address the legitimacy dilemma posed by the AI Act, it has in turn created a further legitimacy dilemma. Here, we contend that the Digital Omnibus on AI has prioritised political and operational rationalities over certain legal and cultural rationalities.

## Section 1: The Challenges of Legislative and Regulatory Legitimacy

Legitimacy is central to the functioning of law and regulation as it creates a stable base of acceptance, support and compliance with rules.[12] However, despite its importance, legitimacy is always problematic as it derives from a multitude of rationales that are potentially conflicting and competing with one another. The dilemma faced by legislators and regulators is how to respond to, accommodate and balance these different and potentially conflicting rationales for legitimacy. These concerns cut across Wahlgren's theory of legislative legitimacy and the rich body of scholarship that examines the legitimacy of regulation.[13] In what follows, we bring these two strands of scholarship together to set out how we conceptualise legitimacy.

---

[8] Jens Steffek, 'The Legitimation of International Governance: A Discourse Approach' (2003) 9 European Journal of International Relations 249, 251; Julia Black, 'Constructing and Contesting Legitimacy and Accountability in Polycentric Regulatory Regimes' (2008) 2 Regulation & Governance 137.
[9] Nathalie A Smuha, 'From a "Race to AI" to a "Race to AI Regulation": Regulatory Competition for Artificial Intelligence' (2021) 13 Law, Innovation and Technology 57.
[10] Gary E Marchant, Braden R Allenby and Joseph R Herkert (eds), *The Growing Gap Between Emerging Technologies and Legal-Ethical Oversight: The Pacing Problem* (Springer Netherlands 2011).
[11] Peter Wahlgren, 'The Legitimacy Sphere: Between Law, Culture, Politics and Enforceability' (2010) 56 Scandinavian Studies in Law 428.
[12] Tom R Tyler, 'Psychological Perspectives on Legitimacy and Legitimation' (2006) 57 Annual Review of Psychology 375 <https://doi.org/10.1146/annurev.psych.57.102904.190038>.
[13] Peter Wahlgren, 'The Legitimacy Sphere: Between Law, Culture, Politics and Enforceability' (2010) 56 Scandinavian Studies in Law 428.

Whilst legitimacy 'is an opaque concept on the border between empirical and normative social science,' Wahlgren's understanding of legislative legitimacy sits comfortably between both normative and sociological approaches to the concept.[14] Normative approaches to legitimacy seek to assess acceptability with reference to a set of predetermined 'standards and criteria of legitimacy'.[15] By contrast, sociological approaches to legitimacy, which have their genesis in the writings of Weber, focus on the alignment between a particular system and the broadly held beliefs at the societal level.[16] According to Wahlgren, we can 'decompose the concept of legitimacy into various aspects of rationality which can be analysed separately.'[17] More specifically Wahlgren sets out a framework of legislative legitimacy that rests upon five rationalities, namely political, legal, cultural, operational and internal rationalities.[18]

*Political rationality* recognises that legislative and regulatory design is inherently political. Law and regulation operate within democratic governance and are often the primary tools available to political decision-makers to address societal problems and implement reforms.[19] Here, law and regulation serve as steering instruments to realise specific policy objectives. Whilst political actors may prioritise efficiency, speed and flexibility to amend rules to ensure that policy goals can be frictionlessly achieved, political rationality also requires attention to legislative and regulatory effectiveness to ensure the achievement of intended objectives.[20]

Although legislation and regulation are political instruments, they must be acceptable 'from the point of view of the legal sector', align with legal principles and provide predictability and coherence within the broader normative order.[21] *Legal (formal) rationality* speaks to the upholding of rule-of-law requirements, procedural safeguards, constitutional requirements and stability of the legal order, all of which limit how legislation and regulation is developed and implemented.[22]

Law and regulation should also 'understandable in [their] social context' and accepted by those they affect.[23] *Cultural rationality* demands that law and regulation align with prevailing social norms, values and expectations.[24] In addition, cultural rationality concerns the processes through which law and regulation are developed and communicated. The extent to which those affected by law and regulation are able participate, engage with and be part of the norm development process and how they understand and perceive norms are central to their acceptance.[25] Without this kind of cultural acceptance, people may distrust, resist or simply fail to follow law and regulation.[26]

[14] Jens Steffek, 'The Legitimation of International Governance: A Discourse Approach' (2003) 9 European Journal of International Relations 249, 251.
[15] Id. at 253.
[16] Max Weber, *Economy and Society: Volume I and II* (Guenther Roth and Claus Wittich eds, University of California Press 1922).
[17] Peter Wahlgren, 'The Legitimacy Sphere: Between Law, Culture, Politics and Enforceability' (2010) 56 Scandinavian Studies in Law 428.
[18] Id.
[19] Peter Wahlgren, Lagstiftning: Rationalitet, Teknik, Möjligheter (Jure Förlag 2014), 44.
[20] Id. at 45-46; for more on effectiveness see Maria Mousmouti, Making Legislative Effectiveness an Operational Concept: Unfolding the Effectiveness Test as a Conceptual Tool for Lawmaking, 9 European Journal of Risk Regulation 445 (2018).
[21] Peter Wahlgren, The Legitimacy Sphere: Between Law, Culture, Politics and Enforceability, in: ICT: Legal Issues (ed) Peter Wahlgren (Stockholm: Stockholm Institute for Scandinavian Law, 2010) 427-442, 428.
[22] Peter Wahlgren, Lagstiftning: Rationalitet, Teknik, Möjligheter (Jure Förlag 2014), 46-48.
[23] Peter Wahlgren, The Legitimacy Sphere: Between Law, Culture, Politics and Enforceability, in: ICT: Legal Issues (ed) Peter Wahlgren (Stockholm: Stockholm Institute for Scandinavian Law, 2010) 427-442, 428
[24] Peter Wahlgren, Lagstiftning: Rationalitet, Teknik, Möjligheter (Jure Förlag 2014), 48.

Closely linked, *operational rationality* requires that law and regulation be workable and capable of effective implementation in practice.[27] Effective implementation requires adequate resources, administrative procedures, and oversight mechanisms to ensure that rules can be upheld and applied. At the same time, the practical conditions of different regulatory fields must be taken into account, since the level of supervision, control and institutional support required to enforce legislation and regulation varies depending on the nature and objectives of the regulation.[28]

Finally, *internal rationality* refers to the internal coherence, clarity, consistency and logical structure of the legal and regulatory rules.[29] The internal rationality of legal and regulatory rules is central to their functioning, understandability, implementation and ultimately their acceptance.[30] Whilst this framework sets out predetermined 'standards and criteria of legitimacy' at its heart it is also a concern with congruence between legislation, political objectives and broadly held societal beliefs, expectations and interests.[31]

Although each rationality represents 'a necessary component of legitimacy,' Wahlgren argues that [l]egitimacy is a phenomena created by the balancing of a number of claims of different origin.[32] As Wahlgren notes:

> [T]he sphere of legitimacy is a model underlining the necessity to take all types of rationality into consideration when a new piece of legislation is prepared. The model also emphasizes the necessity to balance the various claims, as the violation of any aspect of rationality may give rise to criticism and indicate that the proposal ought to be redesigned.[33]

An important observation and assumption that Wahlgren makes, however, is that 'the various aspects of rationality often contradict each other.'[34] Moreover, 'aspects of political, legal, cultural, functional and internal rationality raise demands which may be of a conflicting nature' and that a failure to balance competing legitimacy claims may lead to a legitimacy dilemma.[35]

This observation aligns with much contemporary work done on the legitimacy of regulators. Black, for example, contends that '[t]he demands of legitimacy communities may well be directly opposed – to satisfy one will necessarily lead to dissatisfaction of the other.'[36] Here, legitimacy demands based on values of democracy such as representativeness and participation are in a sense incompatible with functionally based legitimacy claims grounded in efficiency and expertise.[37] When faced with potentially competing legitimacy demands, regulators 'face a legitimacy dilemma; what they need to do to be accepted by one part of their environment, within and outside the regulatory regime, is contrary to how they need to respond to another part.'[38] Even where conflicts between legitimacy demands do not create a direct dilemma, they may still burden regulators with competing accountability demands, potentially undermining their effectiveness and survival. [39]

---

[25] Id. at 49–52.
[26] Id. at 48–53.
[27] Peter Wahlgren, The Legitimacy Sphere: Between Law, Culture, Politics and Enforceability, in: ICT: Legal Issues (ed) Peter Wahlgren (Stockholm: Stockholm Institute for Scandinavian Law, 2010) 427-442, 436-437.
[28] Peter Wahlgren, Lagstiftning: Rationalitet, Teknik, Möjligheter (Jure Förlag 2014), 53.
[29] Id. at 54.
[30] Id. at 54-60.
[31] Jens Steffek, 'The Legitimation of International Governance: A Discourse Approach' (2003) 9 European Journal of International Relations 249253; Julia Black, 'Constructing and Contesting Legitimacy and Accountability in Polycentric Regulatory Regimes' (2008) 2 Regulation & Governance 137.
[32] Peter Wahlgren, 'The Legitimacy Sphere: Between Law, Culture, Politics and Enforceability' (2010) 56 Scandinavian Studies in Law 428, 440.
[33] Peter Wahlgren, The Legitimacy Sphere: Between Law, Culture, Politics and Enforceability, in: ICT: Legal Issues (ed) Peter Wahlgren (Stockholm: Stockholm Institute for Scandinavian Law, 2010) 427-442, 442.
[34] Id. at 428.
[35] Id. at 440.

The legitimacy of legislation and regulation rests on a shifting and contingent patchwork of rationales and demands. To build legitimacy in their rules, legislators and regulators must respond to, accommodate and decide how to balance these demands and rationales. While these legitimacy demands and rationales may align, in many cases they conflict and compete with one another. In such situations of competition and conflict between legitimacy rationales and demands, legislators and regulators face a dilemma as they seek to build legitimacy for their rules. These key insights run through both Wahlgren's theory of legislative legitimacy and contemporary scholarship concerned with the legitimacy of regulation. In the following sections, we draw upon these insights to examine the Digital Omnibus on AI.

## Section 2: The Legitimacy of the AI Act and the Dynamics of EU AI Regulation

Building legitimacy for legal and regulatory rules is always a dynamic process. Legitimacy demands and rationales are shaped by and shift with the tides of political, social, economic and technological change. While these changes are sometimes subtle and slow, the AI Act emerged from and exists within rapidly evolving socio-technical landscape. In the following discussion, we contend that three dynamics driven by the evolving socio-technical landscape of AI have created a legitimacy dilemma for the EU institutions in the context of the AI Act. Although the AI Act is not yet in its infancy, the race for AI regulation, the race for AI dominance and the race for regulatory connection have resulted in a wide range of stakeholders calling into question its operationalisation, its objectives and its substance.

### *The race for AI regulation*

The rapid spread of AI has created growing pressure on governments to determine how AI should be governed. As AI systems become increasingly important for economic growth, public administration, national security and everyday social life, states are not only debating how to regulate AI, but are also competing to shape the rules, standards and institutions that will govern it.[40] This competition has contributed to a race to AI regulation. In this context, regulation extends beyond formal legislation to include technical standards, regulatory institutions, soft law instruments, and broader frameworks through which AI systems are managed and controlled. The race to AI regulation is not simply about regulating a new technology, it is also a struggle over who gets to shape the emerging global order surrounding AI. As Geith et al. observe, 'Shaping the international rules and norms that govern AI is … of considerable interest to states—and who wins and loses in such negotiations will shape how this transformative technology becomes governed, with consequences for politics, economics, and society.'[41]

[36] Julia Black, 'Constructing and Contesting Legitimacy and Accountability in Polycentric Regulatory Regimes' (2008) 2 Regulation & Governance 137, 153.
[37] Id.
[38] Id.
[39] Id. at 153-154.
[40] Nathalie A Smuha, 'From a "Race to AI" to a "Race to AI Regulation": Regulatory Competition for Artificial Intelligence' (2021) 13 Law, Innovation and Technology 57.

The United States (US), China and the EU, which can be understood as competing 'digital empires', stand at the centre of the race for AI regulation.[42] Each possesses the technological, economic and regulatory capacity to shape global AI governance according to its own interests and values. These powers have developed distinct models for governing the digital economy that reflect different ideological commitments regarding the relationship between markets, the state, innovation and individual rights. Bradford argues that the US has largely pursued a market-driven model centered on innovation and private enterprise, China has advanced a state-driven model emphasising centralised control and technological sovereignty, while the European Union has promoted a rights-driven model focused on accountability, safety and fundamental rights protection.[43] As these competing models extend beyond their domestic borders, AI regulation has become characterised by regulatory competition in which states seek not only to regulate AI internally, but also to export their preferred regulatory approaches internationally.

It is within this broader geopolitical context that the EU sought to position itself as a global leader in trustworthy AI regulation through the adoption of the AI Act, which entered into force on 1 August 2024 and follows a staggered implementation schedule. Certain provisions, such as the prohibitions on specific AI practices, obligations related to AI literacy and rules governing general-purpose AI models became applicable in 2025. Other core parts of the framework are scheduled to apply in August 2026 and August 2027. This phased rollout was designed to allow regulators and stakeholders to gain practical experience with the initial rules before the full regulatory regime became operational.

With the adoption of the AI Act, the EU sought not only to regulate AI within the internal market, but also to shape the global norms, standards and expectations surrounding trustworthy AI. This strategy reflected the EU's use of regulatory power to influence global digital governance through the export of its legal standards beyond Europe's borders, especially by leveraging the advantages associated with acting as a first mover in digital regulation. As Bradford explains through the concept of the 'Brussels Effect,' EU digital regulations frequently shape the global practices of technology companies, which often extend compliance with EU rules across their worldwide operations in order to standardise products and services internationally.[44] In this way, the AI Act formed part of a broader effort by the EU to establish its rights-driven regulatory model as an international benchmark for AI governance, particularly at a time when there remained no global consensus regarding how AI should be regulated.

---

[41] Johannes Geith, Magnus Lundgren and Jonas Tallberg, Negotiating the EU AI Act: AI Competitiveness, State Preferences, and Bargaining Success in Johannes Geith, The Global Governance of Artificial Intelligence: Actor Preferences, Bargaining Dynamics, and Institutional Design, Doctoral thesis (Stockholm University, Faculty of Social Sciences, Department of Political Science, 2026).
[42] Anu Bradford, Digital Empires: The Global Battle to Regulate Technology (Oxford University Press, 2023), 6.
[43] Id. at 6-7.
[44] Anu Bradford, The Brussels Effect: How the European Union Rules the World (Oxford University Press, 2020); Anu Bradford, Digital Empires: The Global Battle to Regulate Technology (Oxford University Press, 2023), 28.

However, stakeholder consultations conducted throughout 2025 revealed several implementation challenges that, according to the Commission, could hinder the effective rollout of the Act.[45] In particular, many of the technical standards intended to support compliance had not yet been finalised and key guidance documents from the Commission and the AI Office were still under development. Without these supporting instruments, providers, deployers and other regulated actors faced uncertainty about how to interpret, operationalise and implement the Act's requirements. This situation made compliance with certain obligations, such as those related to high-risk AI systems, difficult in practice.

***The race to AI dominance***

The second dynamic is the race to AI dominance and the competition between states and major technology companies to achieve leadership in AI development. States increasingly understood AI as a strategically important technology that may shape future economic power, scientific advancement, industrial competitiveness and military capabilities. Consequently, states are investing heavily in AI research, computational infrastructure, semiconductor production, data access and talent recruitment to strengthen their position in the global technological landscape.

Although the idea is often framed as an 'AI arms race,' the competition extends far beyond military applications.[46] Instead, the race for AI dominance primarily concerns broader technological capabilities, including 'supply chains, infrastructure, industrial base, strategic industries, scientific capability, and prestige achievements.'[47] The underlying concern of states is that leadership in AI may provide major long-term advantages and that AI leadership will be 'critical for future wealth and power.'[48]

The race to AI dominance is also driven by fears of falling behind competitors. Breakthroughs in AI systems, computational capacity or industrial deployment in one state generates pressure on other states to accelerate their own technological development in response. This pressure creates incentivises the development and deployment of increasingly advanced AI systems across sectors such as defense, healthcare, finance, education, manufacturing and public administration.[49] As Mügge and colleagues describe:

> Jurisdictions worldwide tell strikingly similar stories of geoeconomic dynamics and geopolitical struggles over AI leadership. "We" must advance "our" national economy's position in a fierce "race to AI" by all means, simply because we otherwise risk not only foregoing the socio-economic benefits of innovation but also losing the ability to act autonomously from Big Tech and political adversaries.[50]

The race for AI dominance has called into question the approach of the AI Act that on its face sought to prioritise rights over innovation. During the legislative process, many Member States and EU institutions approached AI governance primarily through 'the lenses of fundamental rights, safety, and democratic protection rather than industrial competitiveness alone.'[51] The European Parliament, for example, emphasised the AI Act's role in protecting fundamental rights and ensuring trustworthy AI systems, while the Commission's original proposal stressed the need to strike a balance between innovation and safety.[52]

[45] European Commission, Digital Package: The European Commission Proposes a New Digital Package to Simplify EU Digital Rules and Boost Innovation, General Questions, <https://digital-strategy.ec.europa.eu/en/faqs/digital-package> accessed 10 March 2026
[46] Allan Dafoe, 'AI Governance: Overview and Theoretical Lenses', in Justin B. Bullock, and others (eds), The Oxford Handbook of AI Governance (Oxford University Press, 2024), 21-44, 36.
[47] Id.
[48] Id.
[49] Id.
[50] Daniel Mügge, Regine Paul and Vali Stan, The AI Matrix: Profits, Power, Politics (Agenda Publishing 2026), 11.

However, in the period following the adoption of the AI Act, industry stakeholders expressed increasing concern that the regulatory framework imposed excessive compliance burdens that could hinder innovation and slow the deployment of AI technologies. These concerns were echoed in public debate and policy discussions, where European regulatory complexity was often portrayed as a potential obstacle to technological development and economic growth.[53] Here, for example, the Draghi Report on European Competitiveness spelled out clearly that:

> As in global AI competition 'winner takes most' dynamics are already prevailing, the EU faces now an unavoidable trade-off between stronger ex ante regulatory safeguards for fundamental rights and product safety, and more regulatory light-handed rules to promote EU investment and innovation.[54]

Only a few months after the AI Act was adopted and put forward as a landmark example of the EU's global regulatory capability, it was recast as an obstacle to Europe's competitiveness in the global technological race. As Geith, Lundgren, and Tallberg observe, 'the ink had hardly dried on the Act before major AI companies, industry lobbies, and innovation-oriented states started to attack the new rules for hurting European competitiveness relative to China and the US.'[55] This rapid shift in political discourse created tensions surrounding the AI Act by exposing growing conflicts between competitiveness and innovation driven policy objectives and the AI Act's original emphasis on fundamental rights protection and trustworthy AI regulation.

### *The race for regulatory connection*

The race for regulatory connection concerns the struggle to ensure that legal and regulatory frameworks remain meaningfully connected to evolving socio-technical realities. In the context of AI, the challenge facing regulators is not simply that AI technologies are developing quickly, but that AI systems are simultaneously transforming social practices, institutional behaviour, economic organisation and everyday human interaction. As Bennett Moses explains, the relevant issue is not merely 'technological change,' but broader 'socio-technical change' in the sense that new technologies generate 'new forms of conduct becoming part of social practice.'[56] From this perspective, the regulatory pressures surrounding the AI Act did not arise solely from advances in AI capabilities themselves, such as generative AI or agentic AI, but from the rapid diffusion of these systems across society and the growing perception that AI was essential to Europe's future prosperity, competitiveness and geopolitical relevance.

---

[51] Johannes Geith, Magnus Lundgren and Jonas Tallberg, Negotiating the EU AI Act: AI Competitiveness, State Preferences, and Bargaining Success in: Johannes Geith, The Global Governance of Artificial Intelligence: Actor Preferences, Bargaining Dynamics, and Institutional Design, Doctoral thesis (Stockholm University, Faculty of Social Sciences, Department of Political Science, 2026).
[52] Id.
[53] *See e.g*. Mario Draghi, A Competitiveness Strategy for Europe, The Future of European Competitiveness Part A (2024), 1-71.
[54] Mario Draghi, 'The Future of European Competitiveness Part B: In-Depth Analysis and Recommendations' 79.
[55] Johannes Geith, Magnus Lundgren and Jonas Tallberg, Negotiating the EU AI Act: AI Competitiveness, State Preferences, and Bargaining Success in: Johannes Geith, The Global Governance of Artificial Intelligence: Actor Preferences, Bargaining Dynamics, and Institutional Design, Doctoral thesis (Stockholm University, Faculty of Social Sciences, Department of Political Science, 2026).

In this respect, the AI Act was perceived as suffering from what Brownsword describes as both descriptive and normative disconnection.[57] Descriptive disconnection emerged because 'the snapshot that the regulators took of a technology and related practices' no longer fully corresponded to technological realities following the development and deployment of foundation models, generative AI systems, and AI agents.[58] However, the AI Act also experienced a deeper normative disconnection as political and economic pressures challenged the original balance struck between innovation and the protection of fundamental rights.[59] As geopolitical competition intensified and concerns regarding European competitiveness grew, disagreements emerged about whether the AI Act imposed excessive regulatory burdens on industry, particularly for SMEs and developers of general-purpose AI systems.

Consequently, the concern was not merely that the law had fallen behind technology in a technical sense, but that socio-technical change had destabilised the underlying regulatory compact itself. Brownsword warns that technological change can create a 'regulatory crisis' because regulatees no longer know where they stand and regulators struggle to maintain legal congruence between the regulation and evolving technological realities.[60] Importantly, however, Brownsword also emphasises that not all forms of disconnection should immediately be resolved through rapid purposive reconnection. In some cases, disconnection is 'productive' because it reveals genuine normative disagreement that requires broader democratic and legislative debate rather than swift technical clarification.

Rapidly evolving socio-technical landscapes are constituted not only by technological change, but also by shifting social, political and economic values and objectives. Such landscapes generate a particular set of legitimacy challenges for legislators and regulators. In the context of the AI Act, we have suggested that these legitimacy challenges are driven by three dynamics: the race for AI regulation; the race for AI dominance and; the race for regulatory connection. The Digital Omnibus on AI seeks to respond to these dynamics and the legitimacy challenges they produce. However, as we argue below, in doing so the Digital Omnibus on AI has created a further legitimacy dilemma for the EU institutions.

## Section 3: The Legitimacy of the Digital Omnibus on AI

On 19 November 2025, the European Commission unveiled its Digital Omnibus package, also known as the EU Digital Simplification Package.[61] The initiative seeks to align the substance and implementation of existing digital legislation with the objectives of innovation, growth and competitiveness. The package is made up of two proposed omnibus laws, namely a Regulation

---

[56] Lyria Bennett Moses, "How to Think about Law, Regulation and Technology: Problems with "Technology" as a Regulatory Target" (2013) 5(1) Law, Innovation and Technology 1-20

[57] Roger Brownsword, 'The Challenge of Regulatory Connection', Rights, Regulation, and the Technological Revolution, 160-184, 166.

[58] Id.

[59] Id.

[60] Id. at 160.

[61] European Commission, Digital Omnibus Regulation Proposal, <https://digital-strategy.ec.europa.eu/en/library/digital-omnibus-regulation-proposal> accessed 12 March 2026

simplifying and consolidating parts of the EU's digital acquis including data, privacy and cyber laws (‘Digital Legislation Omnibus’) and a Regulation on the simplification of the implementation of harmonised rules on artificial intelligence (the ‘Digital Omnibus on AI’).[62] The European Commission has stated that the Digital Omnibus on AI is intended to support a ‘clear, simple and innovation friendly implementation’ of the AI Act in a manner consistent with the AI Continent Action Plan and the Apply AI Strategy.[63]

The Digital Omnibus on AI can be grouped into four categories of amendments. First, the proposal links the application of certain rules to the availability of implementation support, particularly by aligning the application of high-risk AI obligations with the availability of standards and compliance tools.[64] Second, it introduces simplification measures, including extending simplified compliance modalities to small mid-cap companies, reducing documentation and registration burdens, and shifting AI literacy responsibilities toward support and promotion by the Commission and Member States.[65] Third, the proposal seeks to improve the effectiveness of the AI Act’s governance, notably by centralising oversight of general-purpose AI models and certain large platforms within the AI Office.[66] Finally, it introduces measures to support compliance and improve procedures, such as expanding regulatory sandboxes, clarifying the relationship between the AI Act and other EU legislation and facilitating the designation of conformity assessment bodies.[67]

In this section, we argue that the Digital Omnibus on AI can be understood a response to the legitimacy dilemma described above. We analyse how the three dynamics that we outlined previously have shaped the evolving legitimacy of the AI Act. Furthermore, we contend that although the Digital Omnibus on AI seeks to address the legitimacy of the AI Act, a further legitimacy dilemma may be created as the EU institutions contend with the competing political, legal, cultural, operational and internal rationalities that emerge within the rapidly evolving socio-technical landscape of AI development and regulation.

***Political rationality***

At first sight, omnibus legislation can be seen as an effective legislative technique for advancing political priorities and efficiently implementing policy change. Since the start of President Ursula von der Leyen’s second mandate, the Commission has engaged in ambitious simplification agenda aimed at reducing regulatory and administrative burdens for businesses and citizens across the EU.[68] Rather than introducing new legislation, a key aspect of the simplification agenda relies on 'targeted changes' to existing acts, bundled into omnibus packages and which seek to amend interconnected frameworks simultaneously.[69] The Commission contends that its omnibus proposals are intended to address 'similar issues embedded in various legislations' and to generate 'maximum synergies and impacts' by bundling amendments to multiple legislative instruments.[70] In doing so, omnibus proposals are framed as a tool to enhance 'efficiency and coherence' across the regulatory landscape while advancing overarching policy objectives.[71]

[62] European Commission, Proposal for a Regulation of the European Parliament and of the Council amending Regulations (EU) 2016/679, (EU) 2018/1724, (EU) 2018/1725, (EU) 2023/2854 and Directives 2002/58/EC, (EU) 2022/2555 and (EU) 2022/2557 as regards the simplification of the digital legislative framework, and repealing Regulations (EU) 2018/1807, (EU) 2019/1150, (EU) 2022/868, and Directive (EU) 2019/1024 (Digital Omnibus), COM(2025) 837 final, 19 November 2025 < https://eur-lex.europa.eu/legal-content/EN/TXT/?uri=CELEX:52025PC0837>; European Commission, Proposal for a Regulation of the European Parliament and of the Council amending Regulations (EU) 2024/1689 and (EU) 2018/1139 as regards the simplification of the implementation of harmonised rules on artificial intelligence (Digital Omnibus on AI), COM(2025) 836 final, 19 November 2025, <https://eur-lex.europa.eu/legal-content/EN/TXT/?uri=CELEX:52025PC0836> accessed 17 March 2026.
[63] European Commission, Digital Package: The European Commission Proposes a New Digital Package to Simplify EU Digital Rules and Boost Innovation, General Questions, <https://digital-strategy.ec.europa.eu/en/faqs/digital-package> accessed 10 March 2026
[64] Id.
[65] Id.
[66] Id.
[67] Id.

As explained above, part of the urgency to 'simplify' the AI Act undoubtedly stems from the geopolitical framing of AI as a strategic race between the EU, the US and China. The dynamic of the race to AI dominance placed political pressure on the EU institutions to buttress and advance the competitiveness of the EU on the global stage. In particular, the Digital Omnibus on AI recognises the central role of small and medium-sized enterprises (SMEs) advancing innovation and competitiveness, and seeks to facilitate their participation by in the AI ecosystem.[72] These political rationalities, which stem from the dynamic of the race for AI dominance, are the driving force behind the Digital Omnibus on AI.

***Legal (formal) rationality***

Legal (formal) rationality speaks to the upholding of rule-of-law requirements, procedural safeguards, constitutional requirements and the stability and predictability of the legal system. Stability and predictability allow for foreseeability and for legal subjects to anticipate how legal rules will apply to their conduct and plan their actions accordingly. If legislators or regulators revise rules frequently or in unpredictable ways, they undermine the ability of legal subjects to rely on legal and regulatory rules as a stable framework for organising their behaviour. Here, what is striking about the Digital Omnibus on AI is that it revises the AI Act is being before many of its core provisions have entered into application. From the perspective of legal rationality, this early intervention risks undermining the stability that legal actors rely upon when preparing for compliance.

Referring explicitly to the Digital Omnibus on AI, Schwemer and colleagues argue that frequent legislative revisions can 'create challenges for companies, citizens and Member States seeking to remain up to date,' and may also 'complicate the ability of academic analysis and legal commentary to keep pace with regulatory developments.'[73] While technological change

[68] Theodoros Karathanasis, The AI Act: Balancing Implementation Challenges and the EU's Simplification Agenda (May 01, 2025), 1-29, 21-22, available at SSRN: https://ssrn.com/abstract=5311501 or http://dx.doi.org/10.2139/ssrn.5311501
[69] Id. at 22.
[70] European Commission, Simplification: The European Commission Works to Strengthen EU Competitiveness while Protecting Economic, Social, and Environmental Objectives, <https://commission.europa.eu/law/law-making-process/better-regulation/simplification-and-implementation/simplification_en#omnibus-proposals> accessed 10 March 2026.
[71] Id.
[72] Mario Draghi, A Competitiveness Strategy for Europe, The Future of European Competitiveness Part A (2024), 1-71, 8 (stating "…we claim to favour innovation, but we continue to add regulatory burdens onto European companies, which are especially costly for SMEs and self-defeating for those in the digital sectors. More than half of SMEs in Europe flag regulatory obstacles and the administrative burden as their greatest challenge.").
[73] Sebastian Felix Schwemer, Emily Weitzenboeck, Samson Esayas, Arnesen Strøm, Andre Lars, Live Sunniva Hjort, Malcolm Langford, Rebecca Schmidt, Tobias Mahler, Beata Paragi, Malgorzata Agnieszka Cyndecka,

continues to occur at a rapid pace, it is notable that earlier cornerstone instruments, such as the e-Commerce Directive and the General Data Protection Regulation, have remained in force for comparatively long periods.[74] Here, Schwemer and colleagues highlight that maintaining an appropriate balance between regulatory adaptability and legislative stability remains an important consideration in the ongoing development of the EU's digital regulatory framework.[75]

The Digital Omnibus on AI raises further questions concerning transparency and accountability. As scholars have noted in the context of omnibus legislation more broadly, large legislative packages can 'cloud the pathways of representation' and weaken democratic accountability because legislators may highlight certain provisions to justify their vote while minimising others.[76] The Digital Omnibus on AI consolidates numerous amendments to the AI Act into a single legislative proposal that obscures the individual policy choices embedded within the package. The legislative proposal bundles together amendments ranging from changes to compliance timelines and documentation requirements to revisions affecting obligations such as AI literacy and governance arrangements. In so doing, the Digital Omnibus on AI renders it difficult for observers to clearly identify and evaluate the position of political actors on each specific amendment.

Omnibus legislation can 'simplify legislation without altering the purpose, nature, or scope of the underlying rules' only where it is 'properly constrained and transparently executed.'[77] In the case of the Digital Omnibus on AI, however, several of the proposed amendments go beyond 'technical alignment' or 'administrative housekeeping.'[78] Instead, they introduce changes that alter the scope and operation of core regulatory obligations within the AI Act. This raises concerns that the omnibus technique may obscure the substantive importance of the amendments by presenting them as limited adjustments rather than more meaningful policy revisions.[79] For example, the European Data Protection Board (EDPB) and the European Data Protection Supervisor (EDPS) have challenged the characterisation of the omnibus reform as involving only minor or technical amendments.[80]

---

and Riccardo Lazzardi, The Oslo Submission on the Digital Omnibus and the Digital Fitness Check (March 11, 2026), 1-12, 2, available at SSRN: https://ssrn.com/abstract=

[74] Id.

[75] Id.

[76] Glen S. Krutz, Omnibus Legislating in the U.S. Congress, in: Ittai Bar-Siman-Tov (ed.) *Comparative Multidisciplinary Perspectives on Omnibus Legislation* (Springer, 2021), 35-51, 49 in: Ittai Bar-Siman-Tov (ed.) *Comparative Multidisciplinary Perspectives on Omnibus Legislation* (Springer, 2021), 49

[77] Politico, Burning through the Rulebook: Europe's Omnibus Fever (16 December 2025), 1-62, 12 <https://www.politico.eu/wp-content/uploads/2025/12/17/burning-through-the-rulebook-europes-omnibus-fever.pdf> accessed 9 March 2026.

[78] Id.; Sebastian Felix Schwemer, Emily Weitzenboeck, Samson Esayas, Arnesen Strøm, Andre Lars, Live Sunniva Hjort, Malcolm Langford, Rebecca Schmidt, Tobias Mahler, Beata Paragi, Malgorzata Agnieszka Cyndecka, and Riccardo Lazzardi, The Oslo Submission on the Digital Omnibus and the Digital Fitness Check (March 11, 2026), 1-12, 3, available at SSRN: https://ssrn.com/abstract=

[79] Politico, Burning through the Rulebook: Europe's Omnibus Fever (16 December 2025), 1-62, 13 <https://www.politico.eu/wp-content/uploads/2025/12/17/burning-through-the-rulebook-europes-omnibus-fever.pdf> accessed 9 March 2026

[80] European Data Protection Board (EDPB) and European Data Protection Supervisor (EDPS), Joint opinion 1/2026 on the Proposal for a Regulation as regards the simplification of the implementation of harmonised rules on artificial intelligence (Digital Omnibus on AI), adopted 20 January 2026, 1-15, 3, <https://www.edpb.europa.eu/system/files/2026-01/edpb_edps_jointopinion_202601_proposal_ai-omnibus_en.pdf > accessed 17 March 2026

The Digital Omnibus on AI raises even more fundamental questions about whether the core requirements of the EU's Better Regulation framework and larger constitutional order have been respected. Within a democratic system, such as the EU, legislation should emerge from an open, pluralistic and transparent process that is supported by careful parliamentary deliberation, committee scrutiny and fact-finding.[81] However, the urgency to pass the Digital Omnibus on AI has resulted in time limits placed on the relevant EU institutions and actors. These time limits not only constrain the ability of these institutions to scrutinise all the provisions of the proposal, it also removes time for hearing and debate.[82]

When assessed against the requirements of the EU's Better Regulation framework, such compressed legislative processes risk weakening the safeguards that support legal rationality and the resulting quality of law and regulation. The Better Regulation agenda seeks to ensure that EU regulation is developed through transparent and evidence-based procedures, including public consultations, impact assessments and internal review processes intended to test the necessity and proportionality of proposed measures. In the case of the Digital Omnibus on AI, however, several of these procedural safeguards have been curtailed or bypassed. The Digital Omnibus on AI was not preceded by a formal impact assessment, despite the Better Regulation Guidelines generally requiring such assessments for initiatives that may have important economic, societal, or fundamental rights implications.[83] The Commission justified this position by stating that:

> The proposal for the Digital Omnibus is not intended to be accompanied by an impact assessment report, as the adjustments to the corpus of law are technical in their nature, do not modify the spirit of the laws...[84]

Despite this, scholars such as Schwemer and colleagues have pointed to the example of AI literacy to illustrate how the absence of a dedicated impact assessment leaves key regulatory costs and practical implications insufficiently examined, particularly where measures extend beyond purely technical adjustments.[85] In addition, the Commission relied on a small set of limited stakeholder consultations (so called 'Reality Check' meetings). These 'Reality Check' meetings, which were dominated by industry representatives with only minimal civil society participation, raise important concerns about whether the consultation process reflected balanced stakeholder engagement.[86] Finally, the Commission relied on anonymised and non-verifiable estimates of compliance costs presented in these meetings, making it difficult to assess the reliability of the data underpinning the proposed amendments.[87]

---

[81] A. Daniel Oliver-Lalana, Omnibus Legislation in Spain: Between Political Expediency, Doctrinal Condemnation, and Judicial Indulgence, in: Ittai Bar-Siman-Tov (ed.) *Comparative Multidisciplinary Perspectives on Omnibus Legislation* (Springer, 2021), 227-255, 230.

[82] Patricia Popelier, The Practice of Omnibus Laws in Belgium: An Empirical Test, in: Ittai Bar-Siman-Tov (ed.) *Comparative Multidisciplinary Perspectives on Omnibus Legislation* (Springer, 2021), 283-313, 286.

[83] European Commission, Proposal for a Regulation of the European Parliament and of the Council amending Regulations (EU) 2024/1689 and (EU) 2018/1139 as regards the simplification of the implementation of harmonised rules on artificial intelligence (Digital Omnibus on AI), COM(2025) 836 final, 19 November 2025, Explanatory Memorandum, 1-10, 5<https://eur-lex.europa.eu/legal-content/EN/TXT/?uri=CELEX:52025PC0836> accessed 17 March 2026; Laura Lazaro Cabrera and Magdalena Maier, Digital Omnibus Threatens Hard-Won AI Safeguards (Centre for Democracy and Technology Europe, 2025,1-12, 10 <https://cdt.org/wp-content/uploads/2025/12/CDT-Europe-Brief-Digital-Omnibus-Threatens-Hard-Won-AI-Safeguards.pdf> accessed 10 March 2026.

[84] European Commission, Call for Evidence on the Digital Package and Omnibus (2025), Simplification – digital package and omnibus, <https://ec.europa.eu/info/law/better-regulation/have-your-say/initiatives/14855-Simplification-digital-package-and-omnibus_en> accessed 19 March 2026.

[85] Sebastian Felix Schwemer, Emily Weitzenboeck, Samson Esayas, Arnesen Strøm, Andre Lars, Live Sunniva Hjort, Malcolm Langford, Rebecca Schmidt, Tobias Mahler, Beata Paragi, Malgorzata Agnieszka Cyndecka, and Riccardo Lazzardi, The Oslo Submission on the Digital Omnibus and the Digital Fitness Check (March 11, 2026), 1-12, 3, available at SSRN: https://ssrn.com/abstract=

From a broader constitutional perspective, the Digital Omnibus on AI appears inconsistent with several general principles of EU law. The absence of a comprehensive impact assessment and the limited evidentiary basis underlying the proposal sit uneasily with the principle of proportionality under Article 5(4) TEU and the participatory and evidence-based law-making commitments reflected in Article 11 TEU and Article 296 TFEU. [88] While the EU institutions enjoy a broad margin of discretion in crafting legislation, the Court of Justice of the European Union (CJEU) has emphasised that legislative measures must be supported by sufficient information to allow an informed assessment of their necessity and proportionality.[89] Where such information is lacking, the legal rationality and legitimacy of the legislative process itself may be called into question.[90]

***Cultural rationality***

Cultural rationality highlights the importance of ensuring that legal and regulatory rules resonates with the expectations and practices of the actors they seeks to govern. The AI Act attempts to regulate a rapidly evolving technological field that involves diverse stakeholders, including AI providers, deployers, regulatory bodies, end-users and civil society.[91] For the regulatory framework to function effectively, its rules must be perceived not only as intelligible and proportionate, but must also align with broader societal concerns surrounding innovation, risk, and fundamental rights.

The growing emphasis on competitiveness and innovation and the race for AI dominance introduced new actors and legitimacy claims into the regulatory debate. Importantly, the Draghi Report reframed the AI Act less as a landmark fundamental rights instrument and more as a potential obstacle to European technological competitiveness. While political and industry actors may welcome the Digital Omnibus on AI as a necessary step toward de-regulation, simplification and competitiveness, human rights advocates have argued that what is framed as regulatory simplification increasingly amounts re-regulation that weakens of fundamental rights protections.[92]

---

[86] European Commission, Proposal for a Regulation of the European Parliament and of the Council amending Regulations (EU) 2024/1689 and (EU) 2018/1139 as regards the simplification of the implementation of harmonised rules on artificial intelligence (Digital Omnibus on AI), COM(2025) 836 final, 19 November 2025, Explanatory Memorandum, 1-10, 1 <https://eur-lex.europa.eu/legal-content/EN/TXT/?uri=CELEX:52025PC0836> accessed 17 March 2026; Laura Lazaro Cabrera and Magdalena Maier, Digital Omnibus Threatens Hard-Won AI Safeguards (Centre for Democracy and Technology Europe, 2025, 1-12, 9 <https://cdt.org/wp-content/uploads/2025/12/CDT-Europe-Brief-Digital-Omnibus-Threatens-Hard-Won-AI-Safeguards.pdf> accessed 10 March 2026.

[87] Laura Lazaro Cabrera and Magdalena Maier, Digital Omnibus Threatens Hard-Won AI Safeguards (Centre for Democracy and Technology Europe, 2025, 1-12, 10 <https://cdt.org/wp-content/uploads/2025/12/CDT-Europe-Brief-Digital-Omnibus-Threatens-Hard-Won-AI-Safeguards.pdf> accessed 10 March 2026.

[88] Politico, Burning through the Rulebook: Europe's Omnibus Fever (16 December 2025), 1-62, 47 <https://www.politico.eu/wp-content/uploads/2025/12/17/burning-through-the-rulebook-europes-omnibus-fever.pdf> accessed 9 March 2026

[89] See e.g. Case C-482/17, Czech Republic v Parliament and Council ECLI:EU:C:2019:1035 para.78-85.

[90] Politico, Burning through the Rulebook: Europe's Omnibus Fever (16 December 2025), 1-62, 4 <https://www.politico.eu/wp-content/uploads/2025/12/17/burning-through-the-rulebook-europes-omnibus-fever.pdf> accessed 9 March 2026 (stating, "Unless the EU finds ways to reconcile simplification with democratic safeguards, it risks entering a phase in which policymaking outruns the mechanisms that anchor legitimacy.").

[91] Liane Colonna, Towards a Methodological Approach for Understanding Human Oversight Requirements under the AI Act, Human Oversight, European Review of Digital Administration & Law (ERDAL), Special Issue: Towards 'Oversight by Design'? Legal Foundations for Effective Oversight in Automated Public Administration (eds. Lena Enqvist, Hanne Marie Motzfeldt, and Markku Suksi)(2025), 41-55, 47-51.

### *Operational rationality*

From the perspective of operational rationality, the Digital Omnibus on AI clearly prioritises the practical enforceability and manageability of the regulatory framework over other forms of rationality. In particular, the Digital Omnibus on AI responds to the legitimacy dilemma created by the race for AI regulation and reflects a concern that the effectiveness of the AI Act ultimately depends not only on the existence of legal obligations, but also on whether the institutional, technical and administrative conditions necessary to implement those obligations are sufficiently developed. The need for the existence of these implementation mechanisms is especially relevant in areas involving high-risk AI systems, where compliance and enforcement depend heavily on harmonised standards and guidance from supervisory authorities as how to precisely interpret some of the broad legal provisions.

### *Internal rationality*

From the perspective of internal rationality, the legitimacy of the Digital Omnibus on AI depends on whether the consolidated amendments preserve coherence within the broader regulatory framework. Gluck observes that ‘from a statutory interpretation standpoint, omnibus bills pose particular challenges for common doctrinal assumptions regarding legislative perfection.’[93] In practice, the legislative process surrounding omnibus measures may become rushed, politically compromised and procedurally compressed. In so doing, the legislative process may limit opportunities for careful drafting and systematic review. As was seen with the initial drafting of the AI Act, political compromises are often negotiated behind closed doors and late in the legislative process. This resulted in provisions being inserted or modified at the last minute as part of broader political bargains. Under such conditions, errors or linguistic inconsistencies may arise that complicate interpretation and undermine expectations of coherence. Popelier succinctly explains, ‘…the procedures that affect the democratic quality of the law, also affect its substantive or legal quality. Studies, opinions and parliamentary debate do not only make the law more legitimate, they also make sure that the law is well reasoned and balanced.’[94]

In the context of the Digital Omnibus on AI, there was pressure to reach agreement on the proposed amendments before the provisions governing high-risk AI systems became applicable. This compressed timeline created a sense of urgency within the legislative process and likely drove more expedited decision-making than would ordinarily accompany revisions to such a complex regulatory framework. These conditions limit the ability to consider the necessity of amendments and the opportunity for scrutiny, consultation and debate. The accelerated process also limited the ability of external watchdogs, experts and stakeholders to assess the proposals or identify potential inconsistencies prior to adoption.[95] As a result, the urgency surrounding the Digital Omnibus on AI increased the risk that avoidable errors, ambiguities or suboptimal compromises entered the legislative text. Such issues could have been identified and addressed through more extensive parliamentary deliberation and review.

[92] Politico, Burning through the Rulebook: Europe's Omnibus Fever (16 December 2025), 1-62, 16 <https://www.politico.eu/wp-content/uploads/2025/12/17/burning-through-the-rulebook-europes-omnibus-fever.pdf> accessed 9 March 2026.

[93] Abbe R. Gluck, Unorthodox Lawmaking and Legislative Complexity in American Statutory Interpretation in: Ittai Bar-Siman-Tov (ed.) *Comparative Multidisciplinary Perspectives on Omnibus Legislation* (Springer, 2021), 195-225, 201.

[94] Patricia Popelier, The Practice of Omnibus Laws in Belgium: An Empirical Test, in: Ittai Bar-Siman-Tov (ed.) *Comparative Multidisciplinary Perspectives on Omnibus Legislation* (Springer, 2021), 283-313, 287.

As Wahlgren explains, for legislation to be accepted as legitimate it must balance the four different external rationalities: political, legal, cultural and operational.[96] As a result, there is generally only a 'limited sphere of legitimacy.'[97] Furthermore, a legitimate rule should, in addition to being located within the legitimate sphere, also exhibit internal rationality.[98] In our discussion below, we argue that although the Digital Omnibus on AI sought to address a legitimacy dilemma posed by the special dynamics of AI, it has the potential to lead to a further legitimacy dilemma for the European institutions as the reforms to the AI Act simultaneously strengthen and weaken different dimensions of legitimacy.

## Concluding Discussion

According to Krutz, 'omnibus bills are a way to get things done,' especially when the legislative process can be long and cumbersome.[99] The Digital Omnibus on AI provides the EU institutions with a flexible and efficient mechanism for responding to the legitimacy dilemma raised by the AI Act. The Digital Omnibus on AI does so in a way that avoids the delays and institutional burdens associated with reopening the entire legislative framework. In contexts perceived as demanding swift action, legal systems frequently gravitate toward more executive-driven forms of governance that prioritise speed, flexibility and continuous regulatory adjustment over the slower and more deliberative legislative processes traditionally associated with democratic law-making.[100]

While the use of an omnibus helped the EU institutions to reach agreement quickly, it does not mean the outcome is legitimate. Legitimacy requires the balancing of not only political, but also legal, operational and cultural rationalities and demands. The Digital Omnibus on AI compressed public debate and bundled together multiple controversial amendments. It represents a form of fast-tracked law-making, departing from procedural safeguards typically embedded in the EU's legislative process including public consultation under Article 11(3) TEU, impact assessments and inter-service consultations under the Better Regulation framework and the Interinstitutional Agreement on Better Law-Making.[101]

[95] Patricia Popelier, The Practice of Omnibus Laws in Belgium: An Empirical Test, in: Ittai Bar-Siman-Tov (ed.) *Comparative Multidisciplinary Perspectives on Omnibus Legislation* (Springer, 2021), 283-313, 287-288.
[96] Peter Wahlgren, Lagstiftning : Rationalitet, Teknik, Möjligheter (Jure Förlag 2014), 58; Peter Wahlgren, The Legitimacy Sphere: Between Law, Culture, Politics and Enforceability, in: ICT: Legal Issues (ed) Peter Wahlgren (Stockholm: Stockholm Institute for Scandinavian Law, 2010) 427-442, 441.
[97] Id.
[98] Id.
[99] Glen S. Krutz, Omnibus Legislating in the U.S. Congress, in: Ittai Bar-Siman-Tov (ed.) *Comparative Multidisciplinary Perspectives on Omnibus Legislation* (Springer, 2021), 35-51, 36.
[100] Filipe Brito Bastos & Anniek de Ruijter (FNd1), Break or Bend in Case of Emergency?: Rule of Law and State of Emergency in European Public Health Administration, 10 European Journal of Risk Regulation 610, 610–11 (2019)
[101] Interinstitutional Agreement Between the European Parliament, the Council of the European Union and the European Commission on better law-making (13 April 2016), OJ L 123, 12.5.2016, <https://eur-lex.europa.eu/legal-content/EN/TXT/HTML/?uri=CELEX:32016Q0512(01)&from=EN> accessed 10 March 2025; Alberto Alemanno, The Legality of Omnibus Legislation Under EU Law: A Preliminary Analysis of Omnibus I Simplification Directive of CSRD and CSDD and its Legal Consequences on the EU Legal Order

To build legitimacy, regulators often seek to strengthen public trust and acceptance by adopting procedures that enhance transparency, participation, accountability and democratic scrutiny beyond what is strictly required by law. As Black observes, many regulatory bodies have developed extensive systems of consultation, reporting and stakeholder engagement in order to reinforce their legitimacy.[102] In the context of the Digital Omnibus on AI, however, the EU institutions appear to have moved in the opposite direction. Rather than expanding participatory and deliberative mechanisms in response to the profound societal implications of AI regulation, the accelerated use of simplification measures, compressed timelines and limited stakeholder participation and consultation prioritised political and operational rationalities over cultural and legal rationalities. In this respect, the use of the omnibus technique aligned closely with the broader political agenda of the Ursula von der Leyen Commission that has increasingly emphasised innovation, simplification, competitiveness and reducing regulatory burdens in response to the race for AI leadership.[103]

Here, the Digital Omnibus on AI prioritisation of political and operational rationalities reflects a pivot towards output-oriented forms of legitimacy: the legitimacy of the regulatory framework is tied to the EU's ability to generate desirable economic outcomes. Whereas input-oriented legitimacy focuses on participation, representation, and democratic engagement in decision-making processes, output-oriented legitimacy emphasises whether governance institutions are capable of producing effective outcomes and tangible benefits for societal and economic actors.[104] From this perspective, the Digital Omnibus on AI sought to restore confidence in the AI Act by improving its operability, reducing compliance burdens, addressing implementation difficulties and responding to concerns that the original framework risked undermining European competitiveness and innovation capacity. In other words, even as the Omnibus arguably weakened certain forms of input legitimacy through compressed legislative procedures and more selective participation mechanisms, it simultaneously attempted to reinforce the legitimacy of the AI Act by presenting simplification and regulatory adjustment as necessary to ensure effective implementation and the facilitation of innovation and competitiveness.

At the same time, it is important to observe that while improved innovation capacity, regulatory flexibility and economic growth may increase acceptance of the framework among certain political and economic actors, output-based acceptance does not necessarily generate legitimacy within society. Actors who viewed the AI Act primarily as a fundamental rights instrument may not regard competitiveness, simplification or reduced compliance burdens as sufficient justification for weakening of regulatory safeguards. In this regard, legitimacy depends not simply on whether the framework produces economically desirable outcomes, but also on whether it remains faithful to the normative commitments that originally justified the regulation in the first place and aligns with societal expectations. The AI Act was repeatedly presented as a rights-based approach to AI governance that would place human dignity, non-

(November 09, 2025), available at SSRN, 1-33, 4, https://ssrn.com/abstract=5727822 or http://dx.doi.org/10.2139/ssrn.5727822

[102] Julia Black, 'Constructing and Contesting Legitimacy and Accountability in Polycentric Regulatory Regimes' (2008) 2 Regulation & Governance 137.

[103] Theodoros Karathanasis, The AI Act: Balancing Implementation Challenges and the EU's Simplification Agenda (May 01, 2025), 1-29, 22, available at SSRN: https://ssrn.com/abstract=5311501 or http://dx.doi.org/10.2139/ssrn.5311501

[104] Jonas Tallberg, Karin Bäckstrand, and Jan Aart Scholte, 'Introduction: Legitimacy in Global Governance', in Jonas Tallberg, Karin Bäckstrand, and Jan Aart Scholte (eds), Legitimacy in Global Governance: Sources, Processes, and Consequences (Oxford University Press, 2018), 3-19.

discrimination, transparency, and democratic values at the center of AI regulation. To the extent that the Digital Omnibus on AI reframes these protections as obstacles to innovation or administrative burdens requiring simplification, it risks weakening the normative foundations upon which the legitimacy of the AI Act was initially constructed and the cultural rationality of the legislation.

Of course, the Digital Omnibus on AI may illustrate a situation in which EU institutions do not perceive certain legitimacy claims as sufficiently important to alter their regulatory approach. As Black observes, organisations facing multiple and potentially incompatible legitimacy demands may choose to prioritise recognition from some legitimacy communities while disregarding others that are viewed as less essential to achieving political goals or maintaining political support.[105] This dynamic can clearly be seen in the prioritisation of competitiveness, innovation and operational efficiency concerns over the objections raised by civil society organisations, digital rights advocates and human rights groups who have warned that the simplification agenda risks weakening fundamental rights protections. Importantly, while some governance institutions may be capable of ignoring opposition from less powerful societal actors, the legitimacy of EU digital governance has long depended in part upon maintaining support among civil society organisations, academic experts and rights-focused stakeholders who have historically played a key role in shaping European technology regulation. In other words, human rights activities represent and 'important legitimacy-granting audience,' a legitimacy-granting audience that the Digital Omnibus on AI for most part ignores.[106]

We can understand the Digital Omnibus on AI as an attempt at purposive regulatory reconnection that falls somewhere within the sphere of legitimacy. The Digital Omnibus on AI seeks to reconnect the AI Act to technological realities, institutional capacities and broader economic and geopolitical objectives. This regulatory reconnection addresses legitimate forms of unproductive disconnection by attempting to reduce unnecessary complexity and improve regulatory operability. This is particularly true where important elements necessary for implementation, such as harmonised standards, guidelines, codes of practice, and institutional governance structures, were not in place within the timelines originally envisioned by the AI Act. This has resulted in uncertainty regarding compliance, enforcement and practical implementation of the AI Act. At the same time, the Omnibus also treats deeper and more productive forms of disconnection that demand debate and deliberation as though they were merely technical inefficiencies requiring simplification. In doing so, the legislative response clearly prioritises political and operational rationality over legal and cultural rationality.[107]

One of the AI Act's sources of legitimacy was not simply its legal force, but its symbolic claim that democratic societies could govern AI through strong fundamental rights protections while still supporting innovation. Even beyond Europe, the AI Act functioned as evidence that comprehensive AI regulation was both possible and politically justifiable. By framing the AI Act as overly burdensome and in need of rapid simplification even before many of its core obligations had entered into force, the EU institutions have signalled that robust rights-based AI governance is incompatible with competitiveness, innovation and technological development. This is significant as '(f)or many Europeans, safeguarding human dignity, privacy, and non-discrimination is not a negotiable trade-off for technological progress.'[108] This framing thus diminishes the cultural rationality of the AI Act.

[105] Julia Black, 'Constructing and Contesting Legitimacy and Accountability in Polycentric Regulatory Regimes' (2008) 2 Regulation & Governance 137.
[106] Catia Gregoratti and Anders Uhlin, 'Civil Society Protest and the (De)Legitimation of Global Governance Institutions', in Jonas Tallberg, Karin Bäckstrand, and Jan Aart Scholte (eds), Legitimacy in Global Governance: Sources, Processes, and Consequences (Oxford University Press, 2018), 134-150.
[107] Roger Brownsword, 'The Challenge of Regulatory Connection', Rights, Regulation, and the Technological Revolution, 160-184, 166-167.

The legitimacy of the Digital Omnibus on AI depends on whether the political and operational rationality driving the reform remains balanced with the other rationalities that define the sphere of legitimacy. The political and operational motivations underlying the proposal, clearly fall within the political dimension of legitimate law-making. However, legitimacy requires more than responsiveness to political and operational priorities. Law and regulation must also satisfy legal and cultural requirements and demonstrate internal coherence.

The Digital Omnibus on AI may fall within the broader sphere of legitimacy, but only at its margins. The political and operational rationality underlying the Digital Omnibus on AI reflects a broader transformation in EU policymaking in which procedural flexibility and speed are prioritised in order to rapidly realign regulatory frameworks with shifting economic and industrial policy goals.[109] While such responsiveness may be politically and operationally advantageous, it raises concerns about the concentration of agenda-setting power within the executive and the reduced transparency of the policymaking process. [110] To put it differently, the Digital Omnibus on AI may promote political objectives and operational efficiency, but it simultaneously weakened opportunities for broader public deliberation regarding the normative direction of European AI governance and its resonance with broader societal priorities. Legitimacy is not generated solely through political expediency and operational achievements, but through public justification, debate concerning the values and interests underlying the regulatory framework itself and ultimately alignment between regulatory rules and societal expectations.[111]

[108] Nicoletta Rangone, The Paradoxes of the European Union's AI Regulation, https://www.theregreview.org/2026/03/10/rangone-the-paradoxes-of-the-european-unions-ai-regulation/
[109] Politico, Burning through the Rulebook: Europe's Omnibus Fever (16 December 2025), 1-62, 9 <https://www.politico.eu/wp-content/uploads/2025/12/17/burning-through-the-rulebook-europes-omnibus-fever.pdf> accessed 9 March 2026.
[110] Id.
[111] Jens Steffek, 'The Legitimation of International Governance: A Discourse Approach' (2003) 9 European Journal of International Relations 249, 251.